\documentclass[
]{ceurart}

\usepackage{listings}
\usepackage{url}
\begin{document}

\copyrightyear{2024}
\copyrightclause{Copyright for this paper by its authors.
  Use permitted under Creative Commons License Attribution 4.0
  International (CC BY 4.0).}

\conference{BENEVOL24: The 23rd Belgium-Netherlands Software Evolution Workshop, November 21-22, Namur, Belgium}
\title{Development and Adoption of SATD Detection Tools: A State-of-practice Report}

\author[1,2]{Edi Sutoyo}[%
orcid=0000-0002-8413-5070,
email=e.sutoyo@rug.nl,
]
\address[1]{Bernoulli Institute, University of Groningen, Groningen, The Netherlands}
\address[2]{Department of Information Systems, Telkom University, Bandung, Indonesia}

\author[1]{Andrea Capiluppi}[%
orcid=0000-0001-9469-6050,
email=a.capiluppi@rug.nl,
]





\begin{abstract}
Self-Admitted Technical Debt (SATD) refers to instances where developers knowingly introduce suboptimal solutions into code and document them, often through textual artifacts. This paper provides a comprehensive state-of-practice report on the development and adoption of SATD detection tools. Through a systematic review of the available literature and tools, we examined their overall accessibility. Our findings reveal that, although SATD detection tools are crucial for maintaining software quality, many face challenges such as technological obsolescence, poor maintenance, and limited platform compatibility. Only a small number of tools are actively maintained, hindering their widespread adoption. This report discusses common anti-patterns in tool development, proposes corrections, and highlights the need for implementing Findable, Accessible, Interoperable, and Reusable (FAIR) principles and fostering greater collaboration between academia and industry to ensure the sustainability and efficacy of these tools. The insights presented here aim to drive more robust management of technical debt and enhance the reliability of SATD tools.
\end{abstract}

\begin{keywords}
  self-admitted technical debt\sep
  SATD\sep
  detection tools\sep
  FAIR principles\sep
  software engineering practices
\end{keywords}

\maketitle

\section{Introduction}
Technical Debt (TD) broadly refers to suboptimal code or design choices that compromise long-term software maintainability \cite{cunningham1992wycash}. Within this domain, self-admitted technical debt (SATD) refers to instances where developers knowingly document suboptimal code implementations, often through comments in the source code \cite{potdar2014exploratory}. Detecting and managing SATD has become increasingly important due to its significant impact on long-term software maintainability \cite{maldonado2015detecting}. Over the past decade, various tools and approaches have been proposed to automate the detection of SATD, reflecting a growing recognition of its importance to software maintainability. However, despite theoretical advancements, the transition from proposed solutions to widely accessible, practically implementable tools remains a significant challenge in the field.

Several studies have developed automated approaches to identify SATD in source code comments, with tools such as SATDBailiff \cite{AlOmar2022SATDBailiff} and DebtHunter \cite{sala2021debthunter} enabling more effective tracking and management of SATD instances. These tools are critical for managing technical debt, as SATD is prevalent in software projects, affecting 2.4\% to 31\% of files. SATD can persist for extended periods, with a median lifespan ranging from 18 to 172 days and, in some cases, surviving for over 1,000 commits \cite{maldonado2017empirical}. The availability of reliable tools facilitates better identification and management of SATD, helping developers address technical debt and improve overall software quality.

This paper presents a state-of-practice report on the current landscape of SATD detection tools. Based on a recently completed systematic literature review~\cite{sutoyo2024SATDReview}, it provides an evaluation of the available software tools, assessing their functionality and limitations across various dimensions, including accessibility, platform compatibility, and performance in real-world applications. In addition to examining the practical aspects of these tools, this paper identifies recurring anti-patterns—such as poor maintenance, lack of interoperability with modern platforms, and inadequate documentation—that hinder their broader adoption and effectiveness. Based on this analysis, we propose actionable corrections to these anti-patterns to improve the sustainability, usability, and future-proofing of SATD detection tools. Our findings offer insights into the current gaps in the state-of-practice and suggest practical improvements that can drive more robust and reliable SATD management in software development.

This paper is articulated as follows: Section~\ref{sec:_related} presents the related work and reviews existing research on SATD detection tools, focusing on their development, adoption, and challenges. Section~\ref{sec:_methodology} summarizes the methodology and categorises the tools based on their functionality and availability. Section~\ref{sec:_antipatterns} identifies common anti-patterns in SATD tools availability, while Section~\ref{sec:_combating} proposes actionable strategies for combating these anti-patterns. Finally, Section~\ref{sec:_conclusion} concludes the paper.


\section{Related work}
\label{sec:_related}
Research software plays a crucial role in modern science, but its unavailability or malfunction can have serious repercussions. Lack of access to software and data hinders replication of studies, wastes resources on reinventing existing tools, and limits research opportunities \cite{Graham2021Barely}. Many published papers fail to provide accessible data or documentation on outlier handling, impeding reproducibility \cite{Henrik2014Outliers}. At the same time, insufficient software engineering practices in research can undermine Findable, Accessible, Interoperable, and Reusable (FAIR) principles \cite{Graham2021Barely}, and the absence of source code compromises peer review and may bias subsequent work \cite{J2020Recommendations}. To address these issues, experts have long recommended adopting reproducible research practices, which involve publishing both papers and their computational environments \cite{L2017Would, Peng2011Reproducible}. This approach can serve as a minimum standard for evaluating scientific claims when full independent replication is not feasible, ultimately enhancing the reliability and transparency of computational research \cite{Peng2011Reproducible}.

Researchers have shared tools to identify SATD to facilitate empirical studies and improve software maintenance. These tools, such as SATDBailiff \cite{AlOmar2022SATDBailiff} and SATD Detector \cite{Zhongxin2018SATD}, use text mining and machine learning techniques to automatically detect SATD in source code comments, commit messages, pull requests, and issue trackers \cite{Li2023Automatic}. Studies have shown that SATD is common in software projects, affecting 2.4\%-31\% of files \cite{potdar2014exploratory}, and can be effectively identified using automated approaches with high precision and recall \cite{Huang2017Identifying, Zampetti2017Recommending}. Researchers have also explored specific types of SATD, such as “on-hold" SATD \cite{Maipradit2020Wait}, and investigated the gap between admitted and measured technical debt \cite{Pavlic2022Gap}. These tools and studies contribute to better understanding and managing SATD, improving software quality and maintenance practices.

\section{Methodology}
\label{sec:_methodology}
To evaluate the current state-of-practice of software tools designed to detect SATD, we conducted a systematic literature review (SLR) of academic literature and available tools~\cite{sutoyo2024SATDReview}. The review employed search terms such as: \textit{(“self-admitted technical debt" OR SATD) OR (“technical debt" AND NLP) AND (detect* OR identif* OR predict*) AND (“software engineering" OR “software development")}. We intentionally used a broad search string without specifically including the term “\textit{tool}." This approach was intentional to capture as many SATD detection approaches as possible, ensuring a comprehensive review and minimizing the risk of overlooking relevant studies. 

This study builds upon the SLR that identified 68 papers on SATD detection approaches. While the SLR provided a detailed analysis of these approaches, the focus of this paper is narrower, centering specifically on tools identified through the review. We analyze their accessibility and practical utility, highlighting gaps and proposing actionable solutions to address identified anti-patterns. By presenting these contributions, this paper complements the SLR by emphasizing practical insights that can guide the development and improvement of SATD detection tools.

From these 68 studies, we carefully reviewed each paper to determine which ones not only proposed an approach but also offered a prototype or ready-to-use tool. We attempted to access each tool through the links or repositories provided in the papers. Tools were classified into three categories: 
\begin{itemize}
    \item \textbf{Accessible and functional}: Tools that could be successfully accessed and run.
    \item \textbf{Inaccessible or broken link}: Tools with dead or missing links, making them unavailable.
    \item \textbf{Obsolete or non-functional}: Tools that could be accessed but were incompatible with modern platforms or could not be run successfully.
\end{itemize}

Out of the 68 papers, 60 primarily focused on methodologies, frameworks, or approaches without providing prototypes or implemented tools. These papers enriched our understanding of SATD detection but did not meet the criteria for tool evaluation in this study.

\begin{table}[htpb!]
    \centering
    \caption{SATD detection tools proposed from 2014-2024 - I stands for `Identification', and C for `Categorization'}
    \begin{tabular}{p{1.5cm}p{1.04cm}p{0.5cm}p{0.47cm}p{8.15cm}p{1.65cm}}
        \toprule
        \textbf{Name} & \textbf{Ref.} & \textbf{Year} & \textbf{Task} & \textbf{Description} & \textbf{Category}\\

        \midrule 

        DebtViz & \cite{Li2023Automatic} & 2023 & C & A tool that detects, classifies, visualizes, and monitors SATD, categorizing several debt types on a single platform & Accessible\\

        \midrule
        A browser extension & \cite{Khan2022AutomaticPackages} & 2022 & C & A browser extension using an ML model to automatically classify SATD types in rOpenSci R packages. & Inaccessible\\    

        \midrule
        SATDBailiff & \cite{AlOmar2022SATDBailiff} & 2022 & I & A tool designed to mine, identify, and track SATD & Accessible\\

        \midrule
        FixMe & \cite{phaithoon2021fixme} & 2021 & C & A GitHub bot that is developed to detect, monitor, and notify developers about On-hold SATD in their repositories & Broken link\\
        
        \midrule
        DebtHunter & \cite{sala2021debthunter} & 2021 & I, C & A machine learning-based tool for detecting SATD & Accessible\\

        \midrule   
        SATD Detector & \cite{Zhongxin2018SATD} & 2018 & I & A Java library and Eclipse plug-in that automatically detects SATD in comments and integrates with an IDE for easier management & Obsolete\\

        \midrule       
        eXcomment & \cite{Farias2015AComments, farias2021comment} & 2015 & I & A tool designed to parse Java source code and fetch code comments to identify SATD & Broken link\\



        \bottomrule
    \end{tabular}
    \label{tab:tb_tools}
\end{table}

As shown in Table~\ref{tab:tb_tools}, three tools, namely DebtViz \cite{Li2023Automatic}, SATDBailiff \cite{AlOmar2022SATDBailiff}, and DebtHunter \cite{sala2021debthunter}, are currently accessible. However, two tools, FixMe \cite{phaithoon2021fixme} and eXcomment \cite{Farias2015AComments, farias2021comment}, have broken links. Another tool, described as \enquote{A browser extension} in its paper \cite{Khan2022AutomaticPackages}, does not provide a valid link (inaccessible). Additionally, SATD Detector \cite{Zhongxin2018SATD} is obsolete due to incompatibility with newer environments.

\section{Anti-patterns in SATD tools availability}
\label{sec:_antipatterns}

The state of SATD detection tools presents significant implications for researchers and practitioners. Many tools are outdated or incompatible, limiting the reliability of empirical studies and real-world applications \cite{Costa2018On}. Below, we isolate the implications of the unavailability of SATD detection tools.

\begin{itemize}
\item \textit{Reliability of findings} - A limited number of functional SATD tools restricts the generalizability of experimental results. Using only a few tools can fail to capture the variety of SATD identification techniques, increasing the risk of non-representative findings and affecting reproducibility across different programming languages and projects \cite{Deo2016Prescience}.

\item \textit{Bias in analysis} - SATD tools employ unique algorithms, which may introduce bias in analyses. If only a few tools are used, researchers risk favoring certain types of SATD while neglecting others, limiting comprehensiveness. Access to a diverse range of tools is essential for balanced detection \cite{Algaith2018Finding}.

\item \textit{Incomplete coverage of SATD} - SATD encompasses various debt types, including code, design, and documentation debt \cite{AlOmar2022SATDBailiff}. Each type presents unique challenges and implications for software quality and maintainability. However, many existing SATD detection tools are designed to target specific debt types, often focusing narrowly on code-level debt. This limited scope can result in critical aspects of SATD—such as architectural design flaws or insufficient documentation—being overlooked. Such gaps in detection not only skew research conclusions but also hinder effective debt management practices in real-world projects \cite{Patel2022Making}.

\item \textit{Technological obsolescence} - Outdated tools and broken links highlight issues regarding software maintenance in the SATD detection community. Obsolescence hampers usability and sustainability, making it difficult for practitioners to adopt these tools. Continuous updates and adherence to best practices are crucial for maintaining relevance and accessibility \cite{Costa2018On}. Technological obsolescence is particularly evident in tools like SATD Detector, which became incompatible with modern platforms after its initial release.

To ensure long-term sustainability, it is vital to foster community collaboration in tool development, incorporating FAIR principles to enhance the usability of SATD detection tools \cite{Sonabend2024FAIR}.

\item \textit{Sustainability and reproducibility} - Maintaining SATD tools through academic-industry partnerships can align development with real-world needs. Regular benchmarking and case studies will help sustain tool accuracy and reliability \cite{AlOmar2022SATDBailiff}. By promoting open-source collaboration, the community can mitigate technological risks and enhance the longevity of SATD tools \cite{Hasselbring2020From}.
\end{itemize}

\section{Combating the anti-patterns}
\label{sec:_combating}

The sustainability of SATD detection tools is crucial for both academia and industry. Many tools become unavailable or outdated shortly after publication, hindering long-term technical debt management \cite{Costa2018On}. Below we discuss some practical actions that the SATD community should discuss for the sustainability of its SATD detection tools.

\paragraph{Promoting diverse tools} Promoting a diverse range of SATD detection tools is vital for improving technical debt identification. By leveraging multiple tools that employ varied detection techniques, it becomes possible to minimize the occurrence of false positives and false negatives, thereby improving the reliability of the results. Studies indicate that combining diverse static analysis tools enhances detection coverage while maintaining manageable false alarm rates \cite{Algaith2018Finding}. 
    
Additionally, creating comprehensive, centralized archives for SATD tools is crucial for ensuring their long-term availability and accessibility \cite{Audemard2020SAT}. These archives serve as repositories where tools can be preserved and maintained, preventing them from becoming inaccessible or forgotten over time. Furthermore, developing adaptable tools enhances resilience against obsolescence \cite{Deo2016Prescience}: adaptable tools should evolve alongside changing software environments, ensuring that they remain relevant and effective in detecting SATD even as programming languages, frameworks and methodologies advance.

\paragraph{Implementing FAIR principles} Adherence to FAIR principles is crucial for the sustainability of SATD tools. Guidelines from other fields, such as biomedical research, offer valuable insights into how FAIR principles can be applied \cite{Patel2022Making}. Tools like FAIRshare and OpenEBench can support FAIR compliance in SATD detection tool development \cite{Patel2022Making,del2022FAIRsoft}. The FAIR-USE4OS guidelines extend these principles to include User-Centered, Sustainable, and Equitable aspects, ensuring tools are reusable, reproducible, and equitable \cite{Sonabend2024FAIR}.

Building on these principles, a key strategy involves hosting SATD tools on reliable, long-term repositories. For example, using GitHub\footnote{\url{https://github.com}} and preserving scientific research outputs (including tools) on platforms like Zenodo\footnote{\url{https://zenodo.org}} helps maintain accessibility, sustainability, and adherence to FAIR principles. These platforms not only safeguard the tools from obsolescence, but also enable widespread sharing, thereby maintaining accessibility and sustainability. Features like version control are vital for tracking changes and maintaining historical records of tool development \cite{spinellis2005version}. This level of transparency encourages collaborative improvement and strengthens the reproducibility of results—both crucial aspects of research and software engineering.

Explicit versioning further enhances the traceability and clarity of tools \cite{stirbu2021introducing}, making it easier for developers and researchers to locate and identify specific releases. As a result, replicating experiments, addressing issues, or building upon existing work becomes more straightforward. Documenting how a tool has evolved over time allows users to choose the version that best fits their needs. For instance, GitHub integrates versioning seamlessly with common workflows, ensuring consistency in updates and deployments, while Zenodo assigns Digital Object Identifiers (DOIs) to each version, providing persistent and reliable references \cite{klein2020persistence}.

\paragraph{Open-source practices for SATD tools} Open-source practices are essential for the sustainability and continued evolution of SATD detection tools. By adopting open-source models, developers can encourage active maintenance and community involvement, significantly reducing the risk of tool obsolescence \cite{Hasselbring2020From}. Open-source ecosystems provide a foundation for collective innovation, where developers, researchers, and practitioners can contribute to the evolution of SATD tools.

One of the core benefits of integrating open-source practices with FAIR principles is the enhancement of transparency and collaboration. This transparency facilitates peer review and validation and accelerates innovation by enabling developers to build upon each work. Adopting the FAIR-USE4OS guidelines further strengthens this approach by emphasizing User-Centered, Sustainable, and Equitable aspects \cite{Sonabend2024FAIR}. These guidelines ensure that SATD tools address diverse user needs, support long-term usability, and promote equitable access to software, enhancing their cross-domain relevance and societal impact.
    
A cornerstone of open-source best practices is straightforward and comprehensive documentation. This includes user guides, developer instructions, and metadata detailing the tool's purpose, dependencies, and functionalities. Well-maintained documentation empowers both novice and experienced users to effectively utilize and contribute to the tool's development \cite{L2017Would, Jimenez2017Four, Corpas2019Four}.

Furthermore, adopting licenses that support open-source distribution, such as MIT, GPL, or Apache licenses, is essential \cite{gamblin2021picking}. These licenses clarify usage rights, encourage reuse, and protect intellectual property, fostering trust among users and contributors. Encouraging the use of standards and modular architectures can also improve interoperability and integration with other tools, making SATD detection solutions more versatile and adaptable to various contexts.

\paragraph{Enhancing academia-industry collaboration} Fostering collaboration between academia and industry is crucial for aligning SATD tools with real-world needs and challenges. Academic research often focuses on theoretical advancements and experimental frameworks, while industry seeks practical solutions that can be seamlessly integrated into existing workflows. Bridging this gap ensures that SATD tools address academic research questions and provide tangible benefits to practitioners managing technical debt in live software systems. Tool benchmarking is a critical step in this process, offering a way to validate the effectiveness, scalability, and usability of SATD tools across diverse scenarios. By utilizing real-world datasets and conducting case studies, researchers can demonstrate the practical applicability of their tools, building trust and interest within the industry \cite{AlOmar2022SATDBailiff}. These collaborative efforts also help identify gaps between research innovations and industry requirements that will enable the iterative refinement of tools to better serve both domains \cite{bikard2019collaboration}.

Continuous evaluation through practical use cases is key to ensuring that SATD detection tools remain adaptable and valuable over time \cite{duvall2007continuous}. By integrating these tools into actual software development and maintenance environments, both researchers and industry stakeholders can observe how the tools perform under varying conditions, such as different programming languages, team sizes, or project complexities. This hands-on feedback allows developers to refine features, optimize performance, and improve user experience. Academia and industry partnerships can also lead to the development of shared benchmarks, datasets, and metrics, fostering standardization and comparability across tools. These efforts promote robust and sustainable tool development practices that directly address industry pain points. Ultimately, such collaboration enhances the effectiveness of SATD detection tools and contributes to improving software quality, reducing technical debt, and fostering innovation in software engineering.
    

\begin{table}[htpb!]
    \centering
    \caption{Mapping between anti-patterns, practical actions and tools affected}
    \begin{tabular}{p{4.1cm}p{5.5cm}p{5cm}}
        \toprule
        \textbf{Anti-pattern} & \textbf{Practical Action} & \textbf{Tool Affected} \\

        \midrule 

        Reliability of findings & Promoting diverse tools & DebtViz, A browser extension, SATDBailiff, FixMe, DebtHunter, SATD Detector, eXcomment \\

        \midrule
        Bias in analysis & Promoting diverse tools &  DebtViz, A browser extension, FixMe, DebtHunter\\    

        \midrule
        Incomplete coverage of SATD & Enhancing academia-industry collaboration & DebtViz, A browser extension, SATDBailiff, FixMe, DebtHunter, SATD Detector, eXcomment \\
        
        \midrule
        Technological obsolescence & Implementing FAIR principles, \hspace{1cm} Open-source practices for SATD tools &  SATD Detector\\    

        \midrule
        Sustainability and reproducibility & Open-source practices for SATD tools  & A browser extension, FixMe, eXcomment \\

        \bottomrule
    \end{tabular}
    \label{tab:tb_mapping}
\end{table}

Table~\ref{tab:tb_mapping} further elaborates on these challenges by mapping anti-patterns to specific practical actions and the tools affected. To combat reliability of findings and bias in analysis, the promotion of diverse tools is emphasized. Tools such as DebtViz, SATDBailiff, FixMe, DebtHunter, SATD Detector, eXcomment, and browser extensions are highlighted as solutions, as leveraging multiple tools reduces inconsistencies and minimizes false positives and negatives. For incomplete coverage of SATD, fostering academia-industry collaboration is proposed to align tool development with real-world needs and ensure comprehensive detection. Tools such as DebtViz, SATDBailiff, FixMe, DebtHunter, SATD Detector, and browser extensions can benefit from this collaboration, resulting in improved validation and effectiveness. To address technological obsolescence, the implementation of FAIR principles and open-source practices is recommended, particularly targeting tools like the SATD Detector. Finally, sustainability and reproducibility can be ensured through open-source practices, which encourage community-driven maintenance and accessibility. Tools such as browser extensions, FixMe, and eXcomment are cited as examples that can benefit from these practices, promoting longevity and reproducibility in SATD detection. 

\section{Conclusion}
\label{sec:_conclusion}
The analysis of SATD detection tools reveals several challenges that hinder their practical adoption and usefulness in research and practice. Many tools suffer from accessibility issues, such as outdated or broken links, reducing their availability for developers and making empirical research more difficult.

To address these issues, the SATD community should adopt strategies that ensure the long-term sustainability and usability of these tools. Applying the FAIR principles can help maintain the relevance of SATD tools and foster stronger collaboration between academia and industry. Such cooperation ensures that tools evolve alongside real-world software development needs and remain accessible for ongoing research.

In addition, open-source practices should be adopted to encourage community-driven maintenance and development, making tools publicly available and encouraging collaborative contributions. Furthermore, regular benchmarking and real-world case studies should be implemented to ensure the relevance and reliability of these tools in diverse settings.

By focusing on sustainability, accessibility, and collaboration, the SATD detection community can create a more diverse and robust ecosystem of tools, better suited to managing technical debt. This will enhance software quality in both academic research and industrial practice, ensuring that SATD remains manageable as software systems grow in complexity.

\bibliography{Main}

\begin{thebibliography}{38}
\expandafter\ifx\csname natexlab\endcsname\relax\def\natexlab#1{#1}\fi
\providecommand{\url}[1]{\texttt{#1}}
\providecommand{\href}[2]{#2}
\providecommand{\path}[1]{#1}
\providecommand{\DOIprefix}{doi:}
\providecommand{\ArXivprefix}{arXiv:}
\providecommand{\URLprefix}{URL: }
\providecommand{\Pubmedprefix}{pmid:}
\providecommand{\doi}[1]{\href{http://dx.doi.org/#1}{\path{#1}}}
\providecommand{\Pubmed}[1]{\href{pmid:#1}{\path{#1}}}
\providecommand{\bibinfo}[2]{#2}
\ifx\xfnm\relax \def\xfnm[#1]{\unskip,\space#1}\fi
\bibitem[{Cunningham(1992)}]{cunningham1992wycash}
\bibinfo{author}{W.~Cunningham},
\newblock \bibinfo{title}{The wycash portfolio management system},
\newblock \bibinfo{journal}{ACM Sigplan Oops Messenger} \bibinfo{volume}{4} (\bibinfo{year}{1992}) \bibinfo{pages}{29--30}.
\bibitem[{Potdar and Shihab(2014)}]{potdar2014exploratory}
\bibinfo{author}{A.~Potdar}, \bibinfo{author}{E.~Shihab},
\newblock \bibinfo{title}{An exploratory study on self-admitted technical debt},
\newblock in: \bibinfo{booktitle}{2014 IEEE International Conference on Software Maintenance and Evolution}, \bibinfo{organization}{IEEE}, \bibinfo{year}{2014}, pp. \bibinfo{pages}{91--100}.
\bibitem[{Maldonado and Shihab(2015)}]{maldonado2015detecting}
\bibinfo{author}{E.~d.~S. Maldonado}, \bibinfo{author}{E.~Shihab},
\newblock \bibinfo{title}{Detecting and quantifying different types of self-admitted technical debt},
\newblock in: \bibinfo{booktitle}{2015 IEEE 7Th international workshop on managing technical debt (MTD)}, \bibinfo{organization}{IEEE}, \bibinfo{year}{2015}, pp. \bibinfo{pages}{9--15}.
\bibitem[{AlOmar et~al.(2022)AlOmar, Christians, Busho, AlKhalid, Ouni, Newman, and Mkaouer}]{AlOmar2022SATDBailiff}
\bibinfo{author}{E.~A. AlOmar}, \bibinfo{author}{B.~Christians}, \bibinfo{author}{M.~Busho}, \bibinfo{author}{A.~H. AlKhalid}, \bibinfo{author}{A.~Ouni}, \bibinfo{author}{C.~Newman}, \bibinfo{author}{M.~W. Mkaouer},
\newblock \bibinfo{title}{Satdbailiff-mining and tracking self-admitted technical debt},
\newblock \bibinfo{journal}{Science of Computer Programming} \bibinfo{volume}{213} (\bibinfo{year}{2022}) \bibinfo{pages}{102693}.
\bibitem[{Sala et~al.(2021)Sala, Tommasel, and Arcelli~Fontana}]{sala2021debthunter}
\bibinfo{author}{I.~Sala}, \bibinfo{author}{A.~Tommasel}, \bibinfo{author}{F.~Arcelli~Fontana},
\newblock \bibinfo{title}{Debthunter: A machine learning-based approach for detecting self-admitted technical debt},
\newblock in: \bibinfo{booktitle}{Proceedings of the 25th International Conference on Evaluation and Assessment in Software Engineering}, \bibinfo{year}{2021}, pp. \bibinfo{pages}{278--283}.
\bibitem[{Maldonado et~al.(2017)Maldonado, Abdalkareem, Shihab, and Serebrenik}]{maldonado2017empirical}
\bibinfo{author}{E.~D.~S. Maldonado}, \bibinfo{author}{R.~Abdalkareem}, \bibinfo{author}{E.~Shihab}, \bibinfo{author}{A.~Serebrenik},
\newblock \bibinfo{title}{An empirical study on the removal of self-admitted technical debt},
\newblock in: \bibinfo{booktitle}{2017 IEEE International Conference on Software Maintenance and Evolution (ICSME)}, \bibinfo{organization}{IEEE}, \bibinfo{year}{2017}, pp. \bibinfo{pages}{238--248}.
\bibitem[{Sutoyo and Capiluppi(2024)}]{sutoyo2024SATDReview}
\bibinfo{author}{E.~Sutoyo}, \bibinfo{author}{A.~Capiluppi},
\newblock \bibinfo{title}{Self-admitted technical debt detection approaches: A decade systematic review},
\newblock \bibinfo{journal}{arXiv preprint arXiv:2312.15020}  (\bibinfo{year}{2024}).
\bibitem[{{Graham Lee} et~al.(2021){Graham Lee}, {Sebastian Bacon}, {Ian Bush}, {L. Fortunato}, {D. Gavaghan}, {T. Lestang}, {Caroline Morton}, {M. Robinson}, {P. Rocca-Serra}, {Susanna-Assunta Sansone}, and {Helena Webb}}]{Graham2021Barely}
\bibinfo{author}{{Graham Lee}}, \bibinfo{author}{{Sebastian Bacon}}, \bibinfo{author}{{Ian Bush}}, \bibinfo{author}{{L. Fortunato}}, \bibinfo{author}{{D. Gavaghan}}, \bibinfo{author}{{T. Lestang}}, \bibinfo{author}{{Caroline Morton}}, \bibinfo{author}{{M. Robinson}}, \bibinfo{author}{{P. Rocca-Serra}}, \bibinfo{author}{{Susanna-Assunta Sansone}}, \bibinfo{author}{{Helena Webb}},
\newblock \bibinfo{title}{Barely sufficient practices in scientific computing},
\newblock \bibinfo{journal}{Patterns}  (\bibinfo{year}{2021}).
\bibitem[{{Henrik Larsson} et~al.(2014){Henrik Larsson}, {Erik Lindqvist}, and {R. Torkar}}]{Henrik2014Outliers}
\bibinfo{author}{{Henrik Larsson}}, \bibinfo{author}{{Erik Lindqvist}}, \bibinfo{author}{{R. Torkar}},
\newblock \bibinfo{title}{Outliers and {Replication} in {Software} {Engineering}},
\newblock \bibinfo{journal}{Asia-Pacific Software Engineering Conference}  (\bibinfo{year}{2014}).
\bibitem[{{J. Brito} et~al.(2020){J. Brito}, {Jun Z Li}, {J. H. Moore}, {C. Greene}, {Nicole A Nogoy}, {L. Garmire}, and {S. Mangul}}]{J2020Recommendations}
\bibinfo{author}{{J. Brito}}, \bibinfo{author}{{Jun Z Li}}, \bibinfo{author}{{J. H. Moore}}, \bibinfo{author}{{C. Greene}}, \bibinfo{author}{{Nicole A Nogoy}}, \bibinfo{author}{{L. Garmire}}, \bibinfo{author}{{S. Mangul}},
\newblock \bibinfo{title}{Recommendations to enhance rigor and reproducibility in biomedical research},
\newblock \bibinfo{journal}{GigaScience}  (\bibinfo{year}{2020}).
\bibitem[{{L. Madeyski} and {B. Kitchenham}(2017)}]{L2017Would}
\bibinfo{author}{{L. Madeyski}}, \bibinfo{author}{{B. Kitchenham}},
\newblock \bibinfo{title}{Would wider adoption of reproducible research be beneficial for empirical software engineering research?},
\newblock \bibinfo{journal}{Journal of Intelligent \& Fuzzy Systems}  (\bibinfo{year}{2017}).
\bibitem[{Peng(2011)}]{Peng2011Reproducible}
\bibinfo{author}{R.~D. Peng},
\newblock \bibinfo{title}{Reproducible {Research} in {Computational} {Science}},
\newblock \bibinfo{journal}{Science} \bibinfo{volume}{334} (\bibinfo{year}{2011}) \bibinfo{pages}{1226--1227}.
\bibitem[{{Zhongxin Liu} et~al.(2018){Zhongxin Liu}, {Qiao Huang}, {Xin Xia}, {Emad Shihab}, {D. Lo}, and {Shanping Li}}]{Zhongxin2018SATD}
\bibinfo{author}{{Zhongxin Liu}}, \bibinfo{author}{{Qiao Huang}}, \bibinfo{author}{{Xin Xia}}, \bibinfo{author}{{Emad Shihab}}, \bibinfo{author}{{D. Lo}}, \bibinfo{author}{{Shanping Li}},
\newblock \bibinfo{title}{Satd {Detector}: A {Text}-{Mining}-{Based} {Self}-{Admitted} {Technical} {Debt} {Detection} {Tool}},
\newblock \bibinfo{journal}{2018 IEEE/ACM 40th International Conference on Software Engineering: Companion (ICSE-Companion)}  (\bibinfo{year}{2018}).
\bibitem[{Li et~al.(2023)Li, Soliman, and Avgeriou}]{Li2023Automatic}
\bibinfo{author}{Y.~Li}, \bibinfo{author}{M.~Soliman}, \bibinfo{author}{P.~Avgeriou},
\newblock \bibinfo{title}{Automatic identification of self-admitted technical debt from four different sources},
\newblock \bibinfo{journal}{Empirical Software Engineering} \bibinfo{volume}{28} (\bibinfo{year}{2023}).
\bibitem[{Huang et~al.(2017)Huang, Shihab, Xia, Lo, and Li}]{Huang2017Identifying}
\bibinfo{author}{Q.~Huang}, \bibinfo{author}{E.~Shihab}, \bibinfo{author}{X.~Xia}, \bibinfo{author}{D.~Lo}, \bibinfo{author}{S.~Li},
\newblock \bibinfo{title}{Identifying self-admitted technical debt in open source projects using text mining},
\newblock \bibinfo{journal}{Empirical Software Engineering} \bibinfo{volume}{23} (\bibinfo{year}{2017}) \bibinfo{pages}{418--451}.
\bibitem[{Zampetti et~al.(2017)Zampetti, Noiseux, Antoniol, Khomh, and Di~Penta}]{Zampetti2017Recommending}
\bibinfo{author}{F.~Zampetti}, \bibinfo{author}{C.~Noiseux}, \bibinfo{author}{G.~Antoniol}, \bibinfo{author}{F.~Khomh}, \bibinfo{author}{M.~Di~Penta},
\newblock \bibinfo{title}{Recommending when {Design} {Technical} {Debt} {Should} be {Self}-{Admitted}},
\newblock in: \bibinfo{booktitle}{2017 {IEEE} {International} {Conference} on {Software} {Maintenance} and {Evolution} ({ICSME})}, volume~\bibinfo{volume}{3}, \bibinfo{organization}{IEEE}, \bibinfo{year}{2017}, pp. \bibinfo{pages}{216--226}.
\bibitem[{Maipradit et~al.(2020)Maipradit, Treude, Hata, and Matsumoto}]{Maipradit2020Wait}
\bibinfo{author}{R.~Maipradit}, \bibinfo{author}{C.~Treude}, \bibinfo{author}{H.~Hata}, \bibinfo{author}{K.~Matsumoto},
\newblock \bibinfo{title}{Wait for it: identifying ``{On}-{Hold}'' self-admitted technical debt},
\newblock \bibinfo{journal}{Empirical Software Engineering} \bibinfo{volume}{25} (\bibinfo{year}{2020}) \bibinfo{pages}{3770--3798}.
\bibitem[{Pavli{\v c} et~al.(2022)Pavli{\v c}, Hli{\v s}, Heri{\v c}ko, and Berani{\v c}}]{Pavlic2022Gap}
\bibinfo{author}{L.~Pavli{\v c}}, \bibinfo{author}{T.~Hli{\v s}}, \bibinfo{author}{M.~Heri{\v c}ko}, \bibinfo{author}{T.~Berani{\v c}},
\newblock \bibinfo{title}{The {Gap} between the {Admitted} and the {Measured} {Technical} {Debt}: An {Empirical} {Study}},
\newblock \bibinfo{journal}{Applied Sciences} \bibinfo{volume}{12} (\bibinfo{year}{2022}) \bibinfo{pages}{7482}.
\bibitem[{Khan and Uddin(2022)}]{Khan2022AutomaticPackages}
\bibinfo{author}{J.~Y. Khan}, \bibinfo{author}{G.~Uddin},
\newblock \bibinfo{title}{{Automatic Detection and Analysis of Technical Debts in Peer-Review Documentation of R Packages}},
\newblock in: \bibinfo{booktitle}{Proceedings - 2022 IEEE International Conference on Software Analysis, Evolution and Reengineering, SANER 2022}, \bibinfo{organization}{\textsc{Institute of Electrical and Electronics Engineers}}, \bibinfo{year}{2022}, pp. \bibinfo{pages}{765--776}. \DOIprefix\doi{10.1109/SANER53432.2022.00094}.
\bibitem[{Phaithoon et~al.(2021)Phaithoon, Wongnil, Pussawong, Choetkiertikul, Ragkhitwetsagul, Sunetnanta, Maipradit, Hata, and Matsumoto}]{phaithoon2021fixme}
\bibinfo{author}{S.~Phaithoon}, \bibinfo{author}{S.~Wongnil}, \bibinfo{author}{P.~Pussawong}, \bibinfo{author}{M.~Choetkiertikul}, \bibinfo{author}{C.~Ragkhitwetsagul}, \bibinfo{author}{T.~Sunetnanta}, \bibinfo{author}{R.~Maipradit}, \bibinfo{author}{H.~Hata}, \bibinfo{author}{K.~Matsumoto},
\newblock \bibinfo{title}{Fixme: A github bot for detecting and monitoring on-hold self-admitted technical debt},
\newblock in: \bibinfo{booktitle}{2021 36th IEEE/ACM International Conference on Automated Software Engineering (ASE)}, \bibinfo{organization}{IEEE}, \bibinfo{year}{2021}, pp. \bibinfo{pages}{1257--1261}.
\bibitem[{Farias et~al.(2015)Farias, Neto, Silva, and Spinola}]{Farias2015AComments}
\bibinfo{author}{M.~A. D.~F. Farias}, \bibinfo{author}{M.~G. D.~M. Neto}, \bibinfo{author}{A.~B.~D. Silva}, \bibinfo{author}{R.~O. Spinola},
\newblock \bibinfo{title}{{A Contextualized Vocabulary Model for identifying technical debt on code comments}},
\newblock in: \bibinfo{booktitle}{2015 IEEE 7th International Workshop on Managing Technical Debt, MTD 2015 - Proceedings}, \bibinfo{organization}{\textsc{Institute of Electrical and Electronics Engineers}}, \bibinfo{publisher}{IEEE}, \bibinfo{year}{2015}, pp. \bibinfo{pages}{25--32}. \DOIprefix\doi{10.1109/MTD.2015.7332621}.
\bibitem[{Farias et~al.(2021)Farias, Mendes, Mendon{\c{c}}a, and Sp{\'\i}nola}]{farias2021comment}
\bibinfo{author}{M.~Farias}, \bibinfo{author}{T.~S. Mendes}, \bibinfo{author}{M.~G. Mendon{\c{c}}a}, \bibinfo{author}{R.~O. Sp{\'\i}nola},
\newblock \bibinfo{title}{On comment patterns that are good indicators of the presence of self-admitted technical debt and those that lead to false positive items.},
\newblock in: \bibinfo{booktitle}{AMCIS}, \bibinfo{year}{2021}.
\bibitem[{Costa et~al.(2018)Costa, Meirelles, and Chavez}]{Costa2018On}
\bibinfo{author}{J.~Costa}, \bibinfo{author}{P.~Meirelles}, \bibinfo{author}{C.~Chavez},
\newblock \bibinfo{title}{On the sustainability of academic software},
\newblock in: \bibinfo{booktitle}{Proceedings of the {XXXII} {Brazilian} {Symposium} on {Software} {Engineering}}, volume~\bibinfo{volume}{6}, \bibinfo{organization}{ACM}, \bibinfo{year}{2018}, pp. \bibinfo{pages}{202--207}.
\bibitem[{Deo et~al.(2016)Deo, Dash, Suarez-Tangil, Vovk, and Cavallaro}]{Deo2016Prescience}
\bibinfo{author}{A.~Deo}, \bibinfo{author}{S.~K. Dash}, \bibinfo{author}{G.~Suarez-Tangil}, \bibinfo{author}{V.~Vovk}, \bibinfo{author}{L.~Cavallaro},
\newblock \bibinfo{title}{Prescience},
\newblock in: \bibinfo{booktitle}{Proceedings of the 2016 {ACM} {Workshop} on {Artificial} {Intelligence} and {Security}}, \bibinfo{organization}{ACM}, \bibinfo{year}{2016}, pp. \bibinfo{pages}{71--82}.
\bibitem[{Algaith et~al.(2018)Algaith, Nunes, Jose, Gashi, and Vieira}]{Algaith2018Finding}
\bibinfo{author}{A.~Algaith}, \bibinfo{author}{P.~Nunes}, \bibinfo{author}{F.~Jose}, \bibinfo{author}{I.~Gashi}, \bibinfo{author}{M.~Vieira},
\newblock \bibinfo{title}{Finding {SQL} {Injection} and {Cross} {Site} {Scripting} {Vulnerabilities} with {Diverse} {Static} {Analysis} {Tools}},
\newblock in: \bibinfo{booktitle}{2018 14th {European} {Dependable} {Computing} {Conference} ({EDCC})}, \bibinfo{organization}{IEEE}, \bibinfo{year}{2018}, pp. \bibinfo{pages}{57--64}.
\bibitem[{Patel et~al.(2022)Patel, Soundarajan, M{\' e}nager, and Hu}]{Patel2022Making}
\bibinfo{author}{B.~Patel}, \bibinfo{author}{S.~Soundarajan}, \bibinfo{author}{H.~M{\' e}nager}, \bibinfo{author}{Z.~Hu},
\newblock \bibinfo{title}{Making {Biomedical} {Research} {Software} {FAIR}: Actionable {Step}-by-step {Guidelines} with a {User}-support {Tool}}  (\bibinfo{year}{2022}).
\bibitem[{Sonabend et~al.(2024)Sonabend, Gruson, Wolansky, Kiragga, and Katz}]{Sonabend2024FAIR}
\bibinfo{author}{R.~Sonabend}, \bibinfo{author}{H.~Gruson}, \bibinfo{author}{L.~Wolansky}, \bibinfo{author}{A.~Kiragga}, \bibinfo{author}{D.~S. Katz},
\newblock \bibinfo{title}{Fair-{USE4OS}: Guidelines for creating impactful open-source software},
\newblock \bibinfo{journal}{PLOS Computational Biology} \bibinfo{volume}{20} (\bibinfo{year}{2024}) \bibinfo{pages}{e1012045}.
\bibitem[{Hasselbring et~al.(2020)Hasselbring, Carr, Hettrick, Packer, and Tiropanis}]{Hasselbring2020From}
\bibinfo{author}{W.~Hasselbring}, \bibinfo{author}{L.~Carr}, \bibinfo{author}{S.~Hettrick}, \bibinfo{author}{H.~Packer}, \bibinfo{author}{T.~Tiropanis},
\newblock \bibinfo{title}{From {FAIR} research data toward {FAIR} and open research software},
\newblock \bibinfo{journal}{it - Information Technology} \bibinfo{volume}{62} (\bibinfo{year}{2020}) \bibinfo{pages}{39--47}.
\bibitem[{Audemard et~al.(2020)Audemard, Paulev{\' e}, and Simon}]{Audemard2020SAT}
\bibinfo{author}{G.~Audemard}, \bibinfo{author}{L.~Paulev{\' e}}, \bibinfo{author}{L.~Simon}, \bibinfo{title}{SAT {Heritage}: A {Community}-{Driven} {Effort} for {Archiving}, {Building} and {Running} {More} {Than} {Thousand} {SAT} {Solvers}}, \bibinfo{publisher}{Springer International Publishing}, \bibinfo{year}{2020}, pp. \bibinfo{pages}{107--113}.
\bibitem[{del Pico et~al.(2022)del Pico, Gelpi, and Capella-Guti{\'e}rrez}]{del2022FAIRsoft}
\bibinfo{author}{E.~M. del Pico}, \bibinfo{author}{J.~L. Gelpi}, \bibinfo{author}{S.~Capella-Guti{\'e}rrez},
\newblock \bibinfo{title}{Fairsoft-a practical implementation of fair principles for research software},
\newblock \bibinfo{journal}{bioRxiv}  (\bibinfo{year}{2022}) \bibinfo{pages}{2022--05}.
\bibitem[{Spinellis(2005)}]{spinellis2005version}
\bibinfo{author}{D.~Spinellis},
\newblock \bibinfo{title}{Version control systems},
\newblock \bibinfo{journal}{IEEE software} \bibinfo{volume}{22} (\bibinfo{year}{2005}) \bibinfo{pages}{108--109}.
\bibitem[{Stirbu and Mikkonen(2021)}]{stirbu2021introducing}
\bibinfo{author}{V.~Stirbu}, \bibinfo{author}{T.~Mikkonen},
\newblock \bibinfo{title}{Introducing traceability in github for medical software development},
\newblock in: \bibinfo{booktitle}{Product-Focused Software Process Improvement: 22nd International Conference, PROFES 2021, Turin, Italy, November 26, 2021, Proceedings 22}, \bibinfo{organization}{Springer}, \bibinfo{year}{2021}, pp. \bibinfo{pages}{152--164}.
\bibitem[{Klein and Balakireva(2020)}]{klein2020persistence}
\bibinfo{author}{M.~Klein}, \bibinfo{author}{L.~Balakireva},
\newblock \bibinfo{title}{On the persistence of persistent identifiers of the scholarly web},
\newblock in: \bibinfo{booktitle}{Digital Libraries for Open Knowledge: 24th International Conference on Theory and Practice of Digital Libraries, TPDL 2020, Lyon, France, August 25--27, 2020, Proceedings 24}, \bibinfo{organization}{Springer}, \bibinfo{year}{2020}, pp. \bibinfo{pages}{102--115}.
\bibitem[{Jim{\' e}nez et~al.(2017)Jim{\' e}nez, Kuzak, Alhamdoosh, Barker, Batut, Borg, Capella-Gutierrez, Chue~Hong, Cook, Corpas, Flannery, Garcia, Gelp{\' i}, Gladman, Goble, Gonz{\' a}lez~Ferreiro, Gonzalez-Beltran, Griffin, Gr{\" u}ning, Hagberg, Holub, Hooft, Ison, Katz, Lesko{\v s}ek, L{\' o}pez~G{\' o}mez, Oliveira, Mellor, Mosbergen, Mulder, Perez-Riverol, Pergl, Pichler, Pope, Sanz, Schneider, Stodden, Suchecki, Svobodov{\' a}~Va{\v r}ekov{\' a}, Talvik, Todorov, Treloar, Tyagi, van Gompel, Vaughan, Via, Wang, Watson-Haigh, and Crouch}]{Jimenez2017Four}
\bibinfo{author}{R.~C. Jim{\' e}nez}, \bibinfo{author}{M.~Kuzak}, \bibinfo{author}{M.~Alhamdoosh}, \bibinfo{author}{M.~Barker}, \bibinfo{author}{B.~Batut}, \bibinfo{author}{M.~Borg}, \bibinfo{author}{S.~Capella-Gutierrez}, \bibinfo{author}{N.~Chue~Hong}, \bibinfo{author}{M.~Cook}, \bibinfo{author}{M.~Corpas}, \bibinfo{author}{M.~Flannery}, \bibinfo{author}{L.~Garcia}, \bibinfo{author}{J.~L. Gelp{\' i}}, \bibinfo{author}{S.~Gladman}, \bibinfo{author}{C.~Goble}, \bibinfo{author}{M.~Gonz{\' a}lez~Ferreiro}, \bibinfo{author}{A.~Gonzalez-Beltran}, \bibinfo{author}{P.~C. Griffin}, \bibinfo{author}{B.~Gr{\" u}ning}, \bibinfo{author}{J.~Hagberg}, \bibinfo{author}{P.~Holub}, \bibinfo{author}{R.~Hooft}, \bibinfo{author}{J.~Ison}, \bibinfo{author}{D.~S. Katz}, \bibinfo{author}{B.~Lesko{\v s}ek}, \bibinfo{author}{F.~L{\' o}pez~G{\' o}mez}, \bibinfo{author}{L.~J. Oliveira}, \bibinfo{author}{D.~Mellor}, \bibinfo{author}{R.~Mosbergen}, \bibinfo{author}{N.~Mulder}, \bibinfo{author}{Y.~Perez-Riverol},
  \bibinfo{author}{R.~Pergl}, \bibinfo{author}{H.~Pichler}, \bibinfo{author}{B.~Pope}, \bibinfo{author}{F.~Sanz}, \bibinfo{author}{M.~V. Schneider}, \bibinfo{author}{V.~Stodden}, \bibinfo{author}{R.~Suchecki}, \bibinfo{author}{R.~Svobodov{\' a}~Va{\v r}ekov{\' a}}, \bibinfo{author}{H.-A. Talvik}, \bibinfo{author}{I.~Todorov}, \bibinfo{author}{A.~Treloar}, \bibinfo{author}{S.~Tyagi}, \bibinfo{author}{M.~van Gompel}, \bibinfo{author}{D.~Vaughan}, \bibinfo{author}{A.~Via}, \bibinfo{author}{X.~Wang}, \bibinfo{author}{N.~S. Watson-Haigh}, \bibinfo{author}{S.~Crouch},
\newblock \bibinfo{title}{Four simple recommendations to encourage best practices in research software},
\newblock \bibinfo{journal}{F1000Research} \bibinfo{volume}{6} (\bibinfo{year}{2017}) \bibinfo{pages}{876}.
\bibitem[{Corpas and Mellor(2019)}]{Corpas2019Four}
\bibinfo{author}{M.~Corpas}, \bibinfo{author}{D.~Mellor},
\newblock \bibinfo{title}{Four simple recommendations to encourage best practices in research software}  (\bibinfo{year}{2019}).
\bibitem[{Gamblin(2021)}]{gamblin2021picking}
\bibinfo{author}{T.~Gamblin}, \bibinfo{title}{Picking an Open Source License at LLNL: Guidance and Recommendations from the Computing Directorate}, \bibinfo{type}{Technical Report}, Lawrence Livermore National Lab.(LLNL), Livermore, CA (United States), \bibinfo{year}{2021}.
\bibitem[{Bikard et~al.(2019)Bikard, Vakili, and Teodoridis}]{bikard2019collaboration}
\bibinfo{author}{M.~Bikard}, \bibinfo{author}{K.~Vakili}, \bibinfo{author}{F.~Teodoridis},
\newblock \bibinfo{title}{When collaboration bridges institutions: The impact of university--industry collaboration on academic productivity},
\newblock \bibinfo{journal}{Organization Science} \bibinfo{volume}{30} (\bibinfo{year}{2019}) \bibinfo{pages}{426--445}.
\bibitem[{Duvall et~al.(2007)Duvall, Matyas, and Glover}]{duvall2007continuous}
\bibinfo{author}{P.~M. Duvall}, \bibinfo{author}{S.~Matyas}, \bibinfo{author}{A.~Glover}, \bibinfo{title}{Continuous integration: improving software quality and reducing risk}, \bibinfo{publisher}{Pearson Education}, \bibinfo{year}{2007}.

\end{thebibliography}

\end{document}